\documentclass[prb,aps,showpacs,floats,floatfix,twocolumn]{revtex4}
\usepackage{color,graphicx,bm,amsmath}

\begin{document}

\newcommand{\note}[1]{\textcolor{red}{#1}}

\newcommand{\pu}{\mbox{$p_{\uparrow}$}}
\newcommand{\pd}{\mbox{$p_{\downarrow}$}}
\newcommand{\su}{\mbox{$s_{\uparrow}$}}
\newcommand{\sd}{\mbox{$s_{\downarrow}$}}
\newcommand{\iu}{\mbox{$i_{\uparrow}$}}
\newcommand{\id}{\mbox{$i_{\downarrow}$}}

\newcommand{\pus}{\mbox{$p_{\uparrow}^*$}}
\newcommand{\pds}{\mbox{$p_{\downarrow}^*$}}
\newcommand{\sus}{\mbox{$s_{\uparrow}^*$}}
\newcommand{\sds}{\mbox{$s_{\downarrow}^*$}}
\newcommand{\ius}{\mbox{$i_{\uparrow}^*$}}
\newcommand{\ids}{\mbox{$i_{\downarrow}^*$}}

\newcommand{\exu}{\mbox{$\phi_{\uparrow}$}}
\newcommand{\exd}{\mbox{$\phi_{\downarrow}$}}

\newcommand{\pusc}{\mbox{$x$}}
\newcommand{\pdsc}{\mbox{$y$}}

\newcommand{\apu}{\mbox{$|\pu|$}}
\newcommand{\apd}{\mbox{$|\pd|$}}
\newcommand{\asu}{\mbox{$|\su|$}}
\newcommand{\asd}{\mbox{$|\sd|$}}
\newcommand{\aiu}{\mbox{$|\iu|$}}
\newcommand{\aid}{\mbox{$|\id|$}}

\newcommand{\gx}{\mbox{$\gamma_x$}}
\newcommand{\gc}{\mbox{$\gamma_c$}}
\newcommand{\gp}{\mbox{$\gamma_p$}}
\newcommand{\gs}{\mbox{$\gamma_s$}}
\newcommand{\gi}{\mbox{$\gamma_i$}}
\newcommand{\g}{\mbox{$\gamma$}}

\newcommand{\op}{\mbox{$\omega_p$}}
\newcommand{\os}{\mbox{$\omega_s$}}
\newcommand{\oi}{\mbox{$\omega_i$}}

\newcommand{\opo}{\mbox{$\omega_p^0$}}
\newcommand{\oso}{\mbox{$\omega_s^0$}}
\newcommand{\oio}{\mbox{$\omega_i^0$}}

\newcommand{\xa}{\mbox{$\mbox{\scriptsize{\it X}}$}}
\newcommand{\ca}{\mbox{$\mbox{\scriptsize{\it C}}$}}
\newcommand{\xap}{\mbox{$\mbox{\scriptsize{\it X}}_p$}}
\newcommand{\xas}{\mbox{$\mbox{\scriptsize{\it X}}_s$}}
\newcommand{\xai}{\mbox{$\mbox{\scriptsize{\it X}}_i$}}
\renewcommand{\cap}{\mbox{$\mbox{\scriptsize{\it C}}_p$}}
\newcommand{\cas}{\mbox{$\mbox{\scriptsize{\it C}}_s$}}
\newcommand{\cai}{\mbox{$\mbox{\scriptsize{\it C}}_i$}}

\newcommand{\xfp}{\mbox{$|\mbox{\scriptsize{\it X}}_p|$}}
\newcommand{\xfs}{\mbox{$|\mbox{\scriptsize{\it X}}_s|$}}
\newcommand{\xfi}{\mbox{$|\mbox{\scriptsize{\it X}}_i|$}}
\newcommand{\cfp}{\mbox{$|\mbox{\scriptsize{\it C}}_p|$}}
\newcommand{\cfs}{\mbox{$|\mbox{\scriptsize{\it C}}_s|$}}
\newcommand{\cfi}{\mbox{$|\mbox{\scriptsize{\it C}}_i|$}}

\newcommand{\ks}{\mbox{$\kappa$}}
\newcommand{\kf}{\mbox{$g_{s}$}}
\newcommand{\kc}{\mbox{$g_{c}$}}
\newcommand{\kp}{\mbox{$g$}}
\newcommand{\kr}{\mbox{$\tilde{g}$}}

\newcommand{\field}{\mbox{$f$}}
\newcommand{\fu}{\mbox{$f_{_{\uparrow}}$}}
\newcommand{\fd}{\mbox{$f_{_{\downarrow}}$}}
\newcommand{\fpu}{\mbox{$f_{\uparrow}$}}
\newcommand{\fpd}{\mbox{$f_{\downarrow}$}}
\newcommand{\afpu}{\mbox{$|f_{\uparrow}|$}}
\newcommand{\afpd}{\mbox{$|f_{\downarrow}|$}}
\newcommand{\afsu}{\mbox{$|f_{s_{\uparrow}}|$}}
\newcommand{\afsd}{\mbox{$|f_{s_{\downarrow}}|$}}

\newcommand{\pol}{\mbox{$\sigma$}}
\newcommand{\polp}{\mbox{$\sigma_p$}}
\newcommand{\pols}{\mbox{$\sigma_s$}}

\newcommand{\ipt}{\mbox{$I_{p}$}}
\newcommand{\ist}{\mbox{$I_{s}$}}

\newcommand{\ipu}{\mbox{$I_{p_{\uparrow}}$}}
\newcommand{\ipd}{\mbox{$I_{p_{\downarrow}}$}}
\newcommand{\isu}{\mbox{$I_{s_{\uparrow}}$}}
\newcommand{\isd}{\mbox{$I_{s_{\downarrow}}$}}
\newcommand{\ith}{\mbox{$I_p^0$}}

\newcommand{\dpp}{\mbox{$\Delta_p$}}
\newcommand{\dps}{\mbox{$\Delta_p^*$}}
\newcommand{\ds}{\mbox{$\Delta_s$}}
\newcommand{\dss}{\mbox{$\Delta_s^*$}}
\newcommand{\di}{\mbox{$\Delta_i$}}
\newcommand{\dis}{\mbox{$\Delta_i^*$}}

\newcommand{\third}{\mbox{$\frac{1}{3}$}}
\newcommand{\half}{\mbox{$\frac{1}{2}$}}
\newcommand{\quarter}{\mbox{$\frac{1}{4}$}}
\newcommand{\twelfth}{\mbox{$\frac{1}{12}$}}
\newcommand{\degree}{\mbox{$^{\circ}$}}
\newcommand{\xthree}{\mbox{$\chi^{(3)}$}}
\newcommand{\xtwo}{\mbox{$\chi^{(2)}$}}
\newcommand{\is}{\mbox{$\rm i$}}

\title{Steady states of a $\chi^3$ parametric oscillator with coupled
polarisations}

\author{P. R. Eastham}

\affiliation{University of Cambridge,
Cavendish Laboratory, Madingley Road, Cambridge CB3 0HE, United Kingdom.}

\author{D. M. Whittaker}

\affiliation{Department of Physics and Astronomy, University of
Sheffield, Sheffield, S3 7RH, United Kingdom.}

\date{\today}

\begin{abstract}
Polarisation effects in the microcavity parametric oscillator are
studied using a simple model in which two $\chi^3$ optical parametric
oscillators are coupled together.  It is found that there are, in
general, a number of steady states of the model under continuous
pumping.  There are both continuous and discontinuous thresholds, at
which new steady-states appear as the driving intensity is increased:
at the continuous thresholds, the new state has zero output intensity,
whereas at the discontinuous threshold it has a finite output
intensity. The discontinuous thresholds have no analog in the
uncoupled device. The coupling also generates rotations of the linear
polarisation of the output compared with the pump, and shifts in the
output frequencies as the driving polarisation or intensity is
varied. For large ratios of the interaction between polarisations to
the interaction within polarisations, of the order of 5, one of the
thresholds has its lowest value when the pump is elliptically
polarised. This is consistent with recent experiments in which the
maximum output was achieved with an elliptically polarised pump.
\end{abstract}

\pacs{42.65.Yj, 71.36.+c, 42.25.Ja, 78.20.Bh}

\maketitle

\section{Introduction}

Semiconductor microcavities are high finesse Fabry-Perot structures,
typically consisting of a planar semiconductor cavity layer bounded by
Bragg mirrors. The mirrors confine two-dimensional photons, which mix
with the exciton states of quantum wells embedded in the cavity. Such
mixing gives a type of two-dimensional polariton known as a ``cavity
polariton''\cite{review1}. The non-linear dynamics of cavity
polaritons corresponding to polariton-polariton scattering has been
extensively studied, due to the possibility of observing bosonic
effects such as stimulated scattering. Recent experiments have
demonstrated new aspects to the non-linear dynamics of \emph{coherent}
polaritons: parametric oscillation and amplification.

Parametric oscillation and amplification in microcavities has been
demonstrated using both pulsed\cite{savvidis2000,erland2001} and
continuous-wave\cite{stevenson2000} excitation. In these experiments,
a laser tuned near to the energy of the lower polariton mode is used
to generate a coherent polariton field in the microcavity. Owing to a
\xthree\/ non-linearity provided by the exciton-exciton interaction,
this pump mode is coupled to ``signal''and ``idler'' modes at lower
and higher energies.  The coupling corresponds to the scattering of
pairs of pump polaritons into the signal and idler, in contrast to the
coupling in the conventional \xtwo\/ optical parametric
oscillator\cite{shen}, which corresponds to the fission of pump
photons. Above a critical pump intensity, the gain due to the
nonlinearity outweighs the damping of the signal and idler modes. When
this occurs, the steady-state in which there is a single coherent
field at the pump becomes unstable towards a state which also has
coherent fields at the signal and idler. In the continuously pumped
experiments, this instability develops spontaneously, and the new
steady-state is reached. In the pulsed experiments of Ref.\ \
\onlinecite{savvidis2000} however, there is not enough time for the
instability to develop spontaneously before the excitation pulse is
over. Instead it is triggered using a second ``seed'' laser pulse
injected into the signal mode.

The theory of microcavity parametric oscillation and amplification was
initially developed by Ciuti {\it et al.}\cite{ciuti2000} and
Whittaker\cite{whittaker2001}. In the former, the pulsed measurements
of Ref.\ \onlinecite{savvidis2000} are treated within a quantum optics
formalism, while the latter used classical nonlinear optics to explain
the steady-state behaviour. The dynamical equations which occur in
both models are the same, demonstrating that the phenomena are
essentially classical effects, with the exception of the incoherent
luminescence which occurs below threshold.\cite{ciuti2001} More recent
theoretical work by Savasta et al.\cite{savasta2003} includes
frequency-dependent nonlinearities and nonlinear absorption, arising
from exciton-exciton correlations.

Parametric oscillation requires a significant electromagnetic response
at the wavevectors and frequencies of the signal and idler. Thus the
signal and idler must lie near the polariton dispersion. The signal
and idler must also satisfy the requirements of wavevector and
frequency conservation in the generation process. In the conventional
parametric oscillator, these two requirements are usually met by
exploiting birefringence\cite{shen}. Thus the polarisations of the
fields for which the device operates are prescribed. In the
microcavity parametric oscillator however, the unusual dispersion of
cavity polaritons allows them to be met irrespective of the
polarisations. This is achieved for pump fields near to a particular
``magic'' wavevector. In this paper we study the effects of these
polarisation degrees-of-freedom.

There are two recent experiments on the polarisation effects in
microcavities that are resonantly pumped near to the magic wavevector,
one using continuous wave excitation\cite{tartakovskii2000}, and one
using pulsed excitation\cite{lagoudakis2002}. Both these papers argue
that their observations imply the existence of interactions between
polaritons of different circular polarisations. While some aspects of
the pulsed data\cite{lagoudakis2002} were recently explained using a
model without such interactions\cite{kavokin2003}, they may still be
necessary to understand the steady-state experiments of Ref.\
\onlinecite{tartakovskii2000}. Furthermore, interactions between
polaritons of different polarisations seem to be necessary to explain
the polarisation dependence of four-wave mixing experiments in
microcavities\cite{kuwata-gonokami1997}. They could originate from
interactions between excitons of different polarisations, which have
been used to explain the polarisation dependence of four-wave mixing
in quantum wells\cite{bott1993,maialle1994}.

Because the parametric oscillator involves the dynamics of coherent
polaritons, and not simply scattering, the consequences of an
interaction between polaritons of different polarisations are not
obvious. In this paper, we investigate the effects of polarisation
coupling on the steady-states of a simple model.

The remainder of this paper is organised as follows. In
section\ \ref{section:model} we present the model. In section\
\ref{section:cw} we explore the steady-states of the model. The
steady-state calculation is subdivided: in section\
\ref{subsec:nopumpdep} we calculate the possible values of the pump
polariton fields, in section\ \ref{subsec:sigidler} we calculate the output
fields for those values of the pump fields, and in section\
\ref{subsec:pumpdepsol} we combine these results with the ``pump
depletion'' equations to determine the behaviour for a particular
external drive. In section\ \ref{section:discussion} we qualitatively
compare our results with the continuously pumped experiments reported
in Ref.\ \onlinecite{tartakovskii2000}, and comment on the
stability of our solutions and possible microscopic origins for our
phenomenological coupling. Finally, section\ \ref{section:conclusions}
summarises our conclusions.

\section{Model}
\label{section:model}

The model we analyse in this paper is a generalisation of the scalar
treatment described in Ref.\ \onlinecite{whittaker2001}. That model
considers the scattering between pump, signal and idler polaritons of
one circular polarisation. The exciton amplitudes in these fields are
denoted by \pu, \su and \iu\/. They are time dependent, so the exciton
field at the pump wavevector ${\rm \bf k_p}$, for example, takes the
form $\pu(t)\exp{\is({\rm \bf k_p}.{\bf r}-\opo t)}$, where \opo\/ is
the lower branch polariton frequency. For simplicity, we assume that
the bare polariton frequencies satisfy the triple resonance condition
$2\omega_p^0=\omega_s^0+\omega_i^0$. We also assume that the time
dependence of the amplitudes is slow compared with the polariton
splitting, so the upper branch can be omitted from the model.  The
scattering is modelled by a term proportional to $|\exu|^4$ in the
Lagrangian density, corresponding to a \xthree\/ nonlinearity. The
equations governing the exciton amplitudes are then
\begin{subequations}
\label{eq:singledynamics}
\begin{eqnarray}
\label{eq:singlepu}
  -\frac{\is}{\xfp^2} \left( \frac{d}{dt} + \gp \right) \pu +2 \ks \su
  \iu \pus &=& \frac{\cap}{\xap} \fpu(t) \\
\label{eq:singlesu}
  -\frac{\is}{\xfs^2}\left( \frac{d}{dt} + \gs \right) \su + \ks \pu^2
  \ius &=& 0 \\
\label{eq:singleiu}
  -\frac{\is}{\xfi^2}\left( \frac{d}{dt} + \gi \right) \iu  	
  + \ks \pu^2 \sus
  &=& 0 
\end{eqnarray}
\end{subequations}
Here \gp\/ etc are the homogeneous line widths of the polariton
states, and \cap\/ and \xap\/ are the amplitudes of the photon and
exciton in the polariton, i.e. the Hopfield coefficients. They appear
because it is only the excitonic part of the polariton which
interacts. $\fpu(t)$ is the external driving field for the pump, in a
frame rotating at the appropriate polariton frequency. For the
continuously pumped situation we consider in the present paper,
$\fpu(t)=f_\uparrow \exp(-i\delta_p t)$, where $\delta_p$ is the pump
detuning. The nonlinear term in the pump equation (\ref{eq:singlepu})
describes the scattering of pairs of polaritons out of the pump field,
while the terms in (\ref{eq:singlesu},\ref{eq:singleiu}) provide the
corresponding growth in the signal and idler fields. The nonlinear
exciton blue-shift\cite{ciuti2000,whittaker2001} is neglected for
simplicity.

To incorporate the polarisation degrees-of-freedom, we introduce pump,
signal and idler fields for the other circular polarisation, denoting
their exciton amplitudes by \pd, \sd and \id\/. Without any coupling
terms, their dynamics is given by the analogs of
(\ref{eq:singledynamics}).  However, we now introduce a second
\xthree\/ excitonic nonlinearity which couples the up and down spin
excitons. Rather than derive a realistic microscopic model of exciton
spin scattering, we treat this phenomenologically by choosing the
simple form $\kp_0 |\exu|^2 |\exd|^2$. We assume that the coefficients
\ks\/ and $\kp_0$ are independent of momentum and energy. While the
former is physically justified because the wavelengths of the
polaritons which scatter are much larger than any excitonic length
scale\cite{ciuti2000}, it is more difficult to rule out the
possibility of an energy dependent process.

The interaction between excitons of opposite spins introduces two new
types of polariton scattering terms into the equations of motion. If
$\kp_0$ is constant as we assume these both have the same strength,
$\kp_0$, but for now we add subscripts in order to distinguish the
processes in Eqs.\ \ref{eq:master}. The first process we describe as
cross-polarisation parametric scattering, where a pair of pump
polaritons of opposite polarisation scatter into a signal and idler
modes, also of opposite polarisation. These terms are written with a
coefficient \kc. The second process we describe as a
polarisation-flip, where a pump and signal, or idler, polariton
exchange polarisations. This flip process is given a strength \kf. A
similar flip process can also occur between signal and idler
polaritons, but this possibility is neglected here, making the
assumption that the signal and idler amplitudes are small compared
with that of the pump.

With these polarisation-coupling terms, the equations for the spin-up
fields are
\begin{widetext}
\begin{subequations}
\label{eq:master}
\begin{eqnarray}
\label{eq:masterpu}
  -\frac{\is}{\xfp^2} \left( \frac{d}{dt} + \gp \right) \pu 
  +2 \ks \su \iu \pus
  + \kc (\sd \iu + \su \id) \pds
  + \kf (\su \sds + \iu \ids) \pd
  &=& \frac{\cap}{\xap} f_\uparrow \exp(-i\delta_p t), \\
\label{eq:mastersu}
  -\frac{\is}{\xfs^2} \left( \frac{d}{dt} + \gs \right) \su 
  + \ks \pu^2 \ius
  + \kc \pu \pd \ids
  + \kf \pu \pds \sd
  &=& 0, \\
\label{eq:masteriu}
  -\frac{\is}{\xfi^2} \left( \frac{d}{dt} + \gi \right) \iu 
  + \ks \pu^2 \sus
  + \kc \pu \pd \sds
  + \kf \pu \pds \id
  &=& 0.
\end{eqnarray}
\end{subequations}
\end{widetext} The equations obeyed by the spin-down fields are given by flipping the
spin labels in Eqs.\ \ref{eq:master}.

In what follows we will not be concerned with the absolute intensities
of the fields or external pumps. This allows us to eliminate the
normal coupling $\ks$ by scaling all the fields and the pumps
according to $\frac{\cap}{\xap} f_\uparrow \to \frac{\cap}{\xap}
f_\uparrow\sqrt{\ks}=F_\uparrow$, and $\pu\to\pu/\sqrt{\ks}$
etc.. After this rescaling, the coupling strengths in
(\ref{eq:master}) are replaced by their ratios to $\ks$:
$\kc\to\kc/\ks=\kp$ etc.

\section{Parametric Oscillation}
\label{section:cw}

For particular amplitudes of the pump fields, equations
\ref{eq:mastersu} and \ref{eq:masteriu} admit harmonic solutions with
finite amplitudes for the signal and idler fields. To find these
steady states, we set $\pu(t)=\pu \exp{(-i \delta_p t)}$ etc. When the
detunings obey
\begin{equation} \label{eq:paraenergycons} 2
\delta_p=\delta_s+\delta_i, \end{equation} the equations for the
signal and idler amplitudes become time-independent; defining complex
rescaled detunings
$\dpp=-(\delta_p^\prime+\is\gp^\prime)=-(\delta_p+\is\gp)/\xfp^2$
etc. they read
\begin{eqnarray} \label{eq:sssig}  \ds \su +  \pu^2 \ius + \kp \pu \pd \ids + \kp \pu
\pds \sd &=& 0, \\ \label{eq:ssidl} \dis \ius +  \pus^2 \su + \kp \pus \pds \sd + \kp
\pus \pd \ids &=& 0.
\end{eqnarray} Along with their spin-flipped counterparts, (\ref{eq:sssig})
and (\ref{eq:ssidl}) form a set of linear homogeneous equations,
parametrised by the pump amplitudes, for the signal and idler
amplitudes: \begin{equation} \label{eq:gensig} M \begin{pmatrix} \su
\\ \ius \\ \sd \\\ids \end{pmatrix}=0. \end{equation} The matrix of
coefficients $M$, combined with the condition
(\ref{eq:paraenergycons}), determines pump amplitudes and detunings
for which steady state operation is possible, and the signal and idler
fields in these steady states, as a function of $\gs,\gi,\delta_p,g$
and the Hopfield coefficients.

\subsection{Allowed Pump Fields and Detunings}
\label{subsec:nopumpdep}

To determine the pump amplitudes and detunings for which steady-state
operation is possible, we note that since (\ref{eq:gensig}) is
homogeneous, solutions with a finite signal and idler are only
possible if $M$ has a zero eigenvalue. In terms of the pump polariton
intensities $I_\uparrow=\apu^2$, $I_\downarrow=\apd^2$, this occurs
when
\begin{eqnarray}
\label{eq:determ}
(\ds \dis-I_\uparrow^2)(\ds \dis-I_\downarrow^2)= \kp^2 I_\uparrow
I_\downarrow \times \\ (2 I_\uparrow - \ds - \dis)(2 I_\downarrow - \ds
- \dis), \nonumber
\end{eqnarray} which directly determines $I_\uparrow$ as a function of
$I_\downarrow$ and the detunings. Since the determined $I_\uparrow$
should be real it also gives an equation, parametrised by
$I_\downarrow$, among the detunings, which combines with
(\ref{eq:paraenergycons}) to determine $\delta_s$ and $\delta_i$ as a
function of $I_\downarrow$.

\begin{figure}[t]
\begin{center}
\includegraphics[width=1\linewidth]{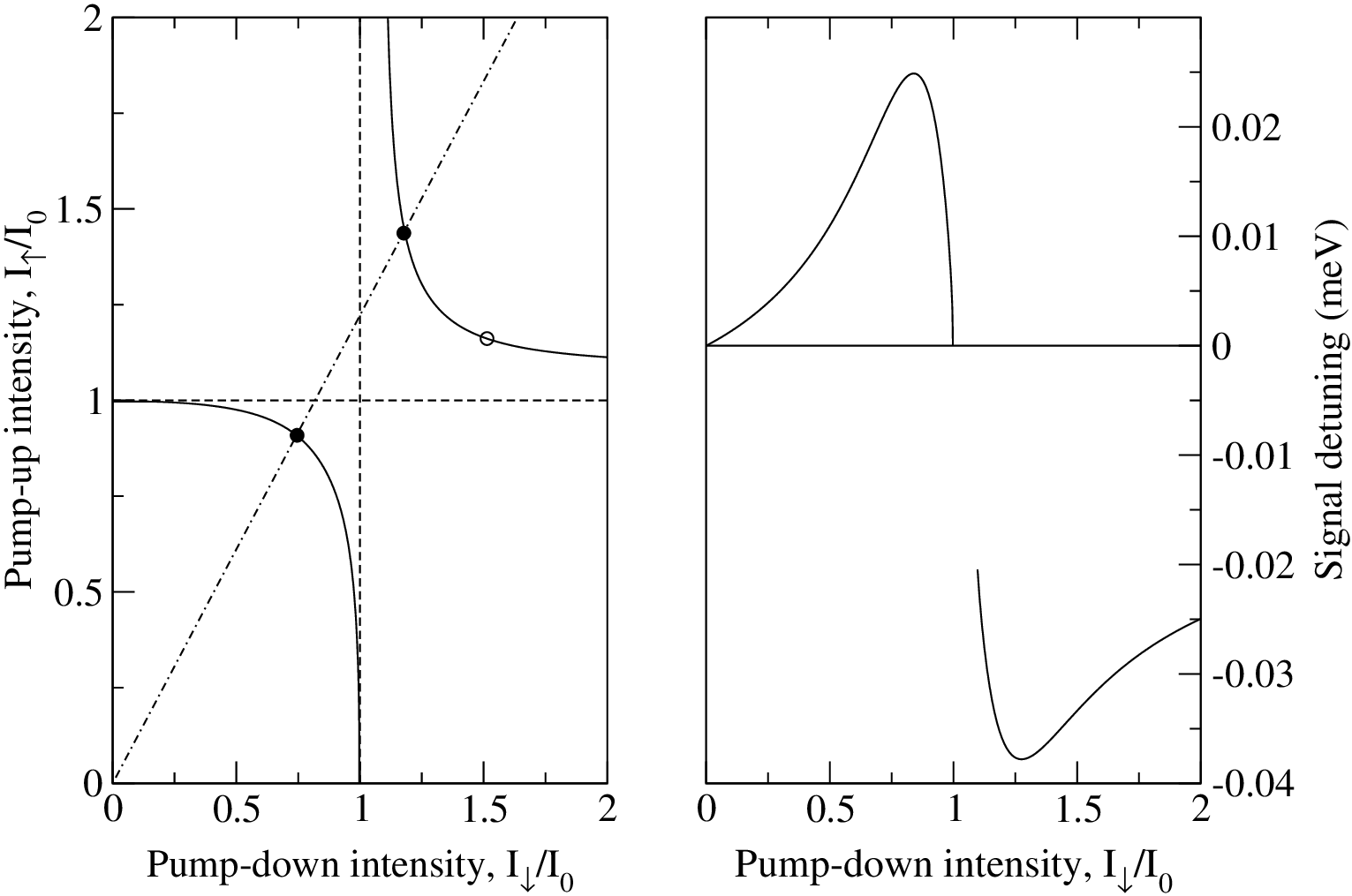}
\end{center}
\caption{Left panel: allowed intensities of the pump fields giving
steady-state operation with $g=0.2$(solid lines), and $g=0$(dashed
lines). The sloping line and dots are for comparison with Fig.\
\ref{fig:pumpdepletionsolution1}. The sloping line is a circularity of
$\sigma=0.1$ for the pump fields, corresponding to the circularity of
the drive used for Fig.\ \ref{fig:pumpdepletionsolution1}. The dots
mark the pump fields at the thresholds of Fig.\
\ref{fig:pumpdepletionsolution1}. Right panel: corresponding signal
detuning for $g=0.2$. For both plots $\gamma_s=0.25
{\mathrm{meV}}$,$\gamma_i=1.0 {\mathrm{meV}}$, $\xfs^2=0.5,
\xfi^2=0.97$, and $\delta_p=0$. The intensities of the pump fields are
given relative to that for steady-state operation with a single
polarisation, $I_0$.}
\label{fig:pumpsolution1}
\end{figure}

In Fig.\ref{fig:pumpsolution1}, we illustrate the pump fields and
signal detunings giving steady-state operation, for a resonant pump
with $\gs=0.25 {\mathrm{meV}} $, $\gi=1 {\mathrm{meV}}$, $g=0.2$, and
the Hopfield coefficients $\xfs^2=0.5$ and $\xfi^2=0.97$. We have
estimated these damping rates and Hopfield coefficients to be those of
the experiment reported in Ref.\ \onlinecite{lagoudakis2002}. The
curves describing the allowed pump fields look very similar to those
for the uncoupled device, shown as dashed lines, except that the
degeneracy, where both polarisations are on threshold, has been
split. The signal detuning shows small deviations from the uncoupled
case, where it would be zero with this resonant pump. These shifts of
the signal detuning from its uncoupled value are due to the spin-flip
processes and the imbalance in the damping of the signal and idler;
without spin-flips or for $\gs^\prime=\gi^\prime$ the condition for
the intensities to be real is $\Im \ds \dis=0$, as in the uncoupled
device.

We can determine whether the form shown in Fig\
\ref{fig:pumpsolution1} is general for small $g$ by using perturbation
theory to calculate how the degeneracy is split by the coupling. We
expand the left-hand side of Eq.\ \ref{eq:determ} to first-order in
the deviation of the pump intensities from the degeneracy and the
change in the detunings compared to the uncoupled case, and take the
right-hand side to be unchanged to leading order. Eliminating the
detunings such that the fields remain real, we find that the shifts in
the pump intensities obey a real quartic form. Owing to its
complexity, we have not studied this quartic in general. However, for
the special case of a resonant pump it becomes
\begin{equation}\label{eq:degeneracysplit} 4I_0^2(a^\prime-4 I_0^2\delta I_\uparrow)(\delta I_\uparrow+\delta
I_\downarrow)^2 +a^{\prime\prime^2}=0,\end{equation} where
$a=a^\prime+i a^{\prime\prime}$ is the right-hand side of Eq.\
\ref{eq:determ} evaluated on the degeneracy. Eq.\
\ref{eq:degeneracysplit} always has the form seen near the degeneracy
in Fig.\ \ref{fig:pumpsolution1}, so the structure of that figure is
general for resonant pumping and small $g$.

We can also use perturbation theory away from the degeneracy, to study
the shift in the allowed pump fields produced by a small $g$. Again
for a resonant pump, the leading deviation in $I_\uparrow$ from an
uncoupled solution in which $I_\uparrow=I_0$ is on threshold while
$I_\downarrow$ is well away from it obeys
\begin{displaymath} \delta
I_\uparrow=-\frac{g^2
I_\downarrow}{2(\gs^\prime\gi^\prime-I_\downarrow^2)}\left(4I_\downarrow\sqrt{\gs^\prime\gi^\prime}-(\gs^\prime-\gi^\prime)^2\right).\end{displaymath}
Thus for a small value of one polarisation, turning on the
cross-polarisation coupling increases the threshold for the other
polarisation. There is therefore a region of $I_\downarrow$, just
above the uncoupled threshold, in which $I_\uparrow(I_\downarrow)$ is
multivalued, although this is not visible on the scale of Fig.\
\ref{fig:pumpsolution1}. Such behaviour is possible because in the
uncoupled device the signal and idler fields of the below-threshold
polarisation are zero, and hence there is no loss through these
channels. With a finite coupling however, these fields become finite,
providing another loss channel. The modes do not always mix in this
way however; for general pump detunings the initial shift can be to
higher or lower thresholds.

\begin{figure}[t]
\begin{center}
\includegraphics[width=1\linewidth]{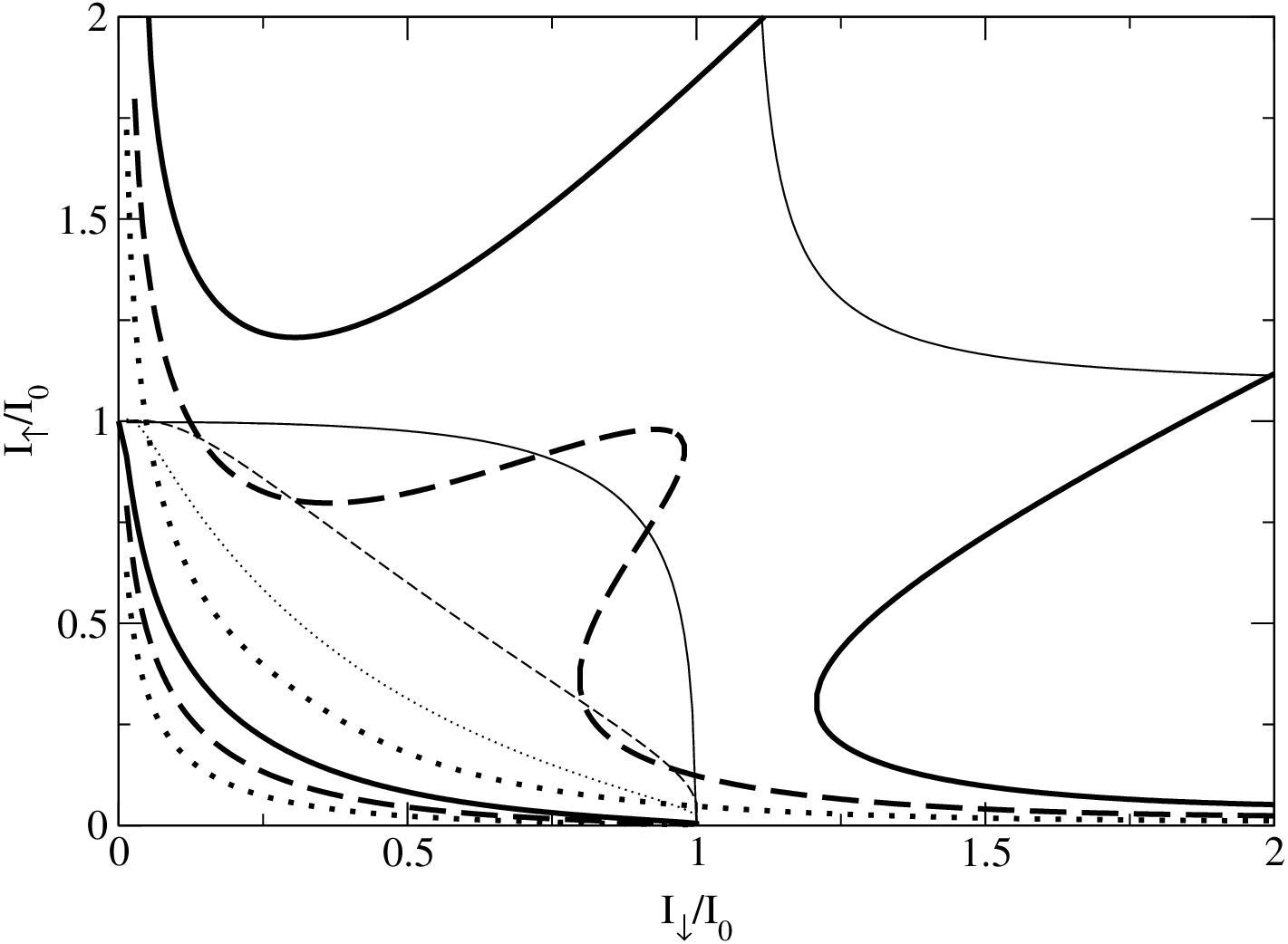}
\end{center}
\caption{Allowed pump-up intensity as a function of pump-down
intensity for $g=0.2$(thin solid lines), 1.0(thin dashed line),
2.0(thin dotted line), 5.0(thick solid lines), 7.0(thick dashed
lines), 10.0(thick dotted lines). The remaining parameters are as used
for figure\ \ref{fig:pumpsolution1}.}
\label{fig:pumpsolution2}
\end{figure}

In the strong-coupling limit, $g\to\infty$, the solutions to Equation\
\ref{eq:determ} for $\gamma_s^\prime\ne\gamma_i^\prime$ have either
$I_\uparrow$ or $I_\downarrow=0$. The approach to this limit is
illustrated in Fig.\ref{fig:pumpsolution2}, where we plot the
allowed pump fields for increasing values of $g$. The lower branch
simply collapses towards the origin. The upper branch disappears, then
reappears as two disjoint branches with asymptotes $I_\uparrow=0$ and
$I_\downarrow=0$. These branches then merge, before finally collapsing
into the origin.

\subsection{Signal and idler fields}
\label{subsec:sigidler}

We now consider the signal and idler fields in the
steady-states. These fields are determined, up to an overall complex
scale factor $z$, by the eigenvector of $M$ corresponding to the zero
eigenvalue.

The eigenvectors of $M$, unlike the eigenvalues, depend on the phases
of the pump fields, $\phi_\uparrow$ and $\phi_\downarrow$. This
dependence can be extracted by noting that $M$ can be written in the
form $S^{\dagger}MS$, where $S$ is a diagonal matrix with entries
$e^{-i\phi_\uparrow},e^{i\phi_\uparrow},e^{-i\phi_\downarrow},e^{i\phi_\downarrow}$.
Thus the phases of the pump fields simply shift the arguments of the
steady-state signal and idler fields: supposing
$\vec{e}=(\su,\ius,\sd,\ids)$ is an eigenvector of $M$ when
$\phi_\uparrow=\phi_\downarrow=0$, then $\vec{e^\prime}=(\su
e^{i\phi_\uparrow},\ius e^{-i\phi_\uparrow},\sd
e^{i\phi_\downarrow},\ids e^{-i\phi_\downarrow})$ is the corresponding
eigenvector for finite phases. The phases of the pump fields have such
a simple effect because there is no phase dependence in the form of
the interaction energy we have chosen. A nonlinearity such as $(\Re
\phi)^4$, rather than $|\phi|^4$, might lead to a more complicated
effect of the pump phases.

\begin{figure}[t]
\begin{center}
\includegraphics[width=1\linewidth]{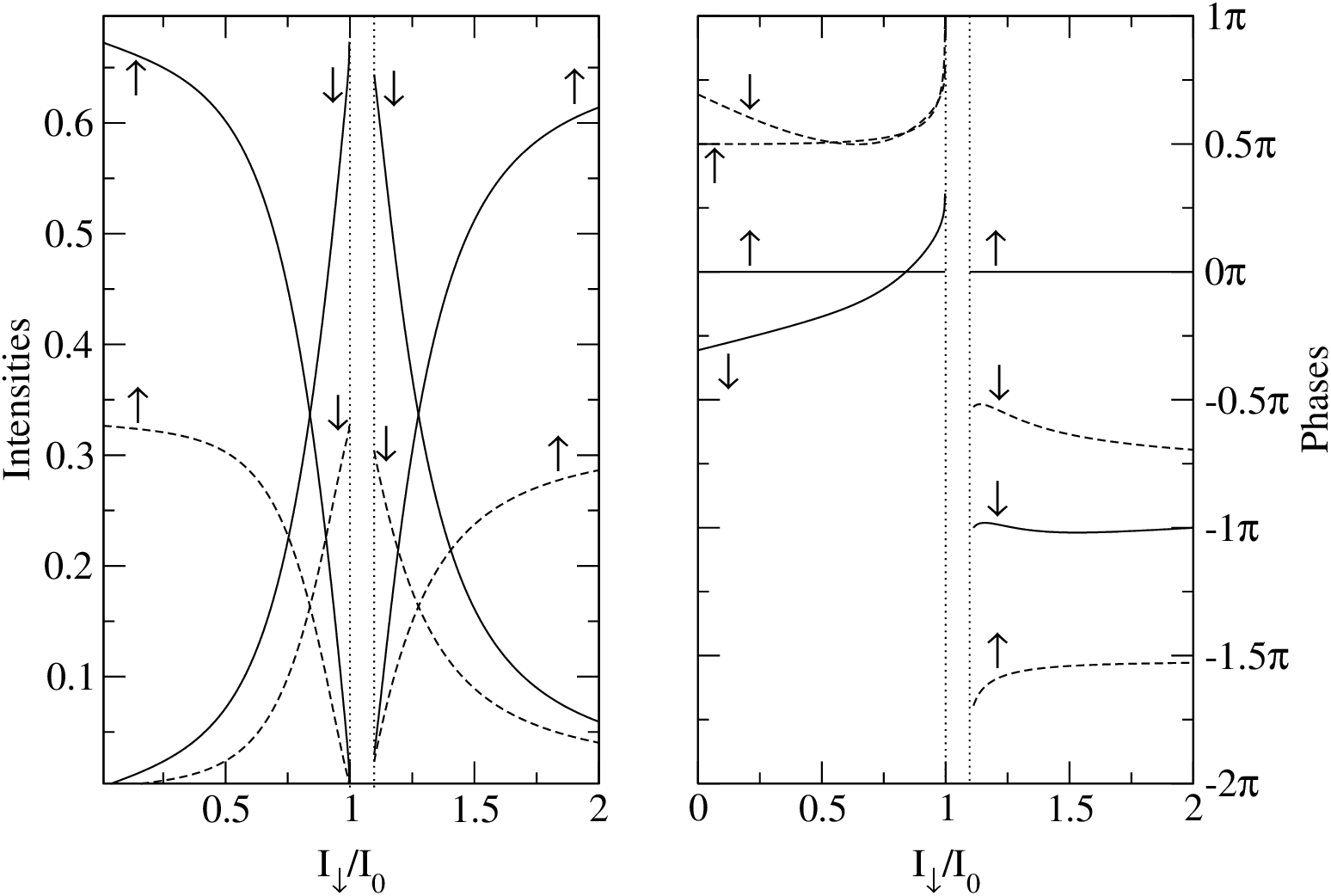}
\end{center}
\caption{Signal fields(solid lines) and conjugate of the idler
fields(dashed lines) corresponding to the steady-states shown in Fig.\
\ref{fig:pumpsolution1}, scaled such that the total intensity is one
and the phase of the signal up component is zero.}
\label{fig:sigidlersolution1}
\end{figure}

In Fig.\ \ref{fig:sigidlersolution1}, we plot the components of the
eigenvector of zero eigenvalue for the pump fields shown in Fig.\
\ref{fig:pumpsolution1}. We have normalised the eigenvector so that
the total intensity is one and the phase of the signal up field is
zero, and taken the pump fields to be real and positive. The idler
fields are always smaller than the corresponding signal fields due to
the stronger damping of the idlers. The crossings of the signal curves
occur when the intensities in the two pump components are equal. In
the region $I_\downarrow/I_0<1$, the polarisation with the largest
pump fields also has the largest signal and idler fields, but that
ordering is reversed in the region $I_\downarrow/I_0>1$. For
$I_\downarrow/I_0<1$, the phase differences between corresponding
components in the two polarisations lie near to zero, while for
$I_\downarrow/I_0>1$ they lie near to $\pi$.

The two circular components of the signal field can be combined to
form, in general, an elliptically polarised state. The phase
differences between the two components of the signal that can be seen
in Fig.\ref{fig:sigidlersolution1} correspond to rotations of the
ellipse describing the signal polarisation compared with that
describing the pump polarisation. Such polarisation rotations are
absent in our model if there are no spin-flip processes.

\subsection{Dependence on the driving fields}
\label{subsec:pumpdepsol}

The steady-state reached in the device is selected from the
possibilities shown in Fig.\ \ref{fig:pumpsolution1} by the external
driving fields, according to the steady-state version of the pump
equation (\ref{eq:masterpu})
\begin{eqnarray}\label{eq:cwpu} \dpp \pu +2 \su \iu \pus + \kp (\sd \iu + \su
\id) \pds \\ + \kp (\su \sds + \iu \ids) \pd = F_\uparrow,
\nonumber\end{eqnarray} and its spin flipped counterpart. The first
term on the left-hand side of Eq.\ \ref{eq:cwpu} describes the bare
response of the pump field, while the remaining ``pump depletion''
terms describe the effect on the pump fields of the nonlinear
processes which generate the signal and idler. The pump equations
(\ref{eq:cwpu}) determine the remaining four real unknowns:
$I_\downarrow$, the arguments of the pump fields, and the total
intensity of the output fields, $|z|^2$. The pump equations are
independent of the overall phase of the zero eigenvector, $\arg z$,
corresponding to a single free phase among the output fields.

To solve the pump equations (\ref{eq:cwpu}), we first extract the
dependence of the phases of the signal and idler on the phases of the
pump, as discussed in section\ \ref{subsec:sigidler}. This gives
\begin{equation}\label{eq:cwp2} e^{i\arg p_\uparrow} \left| L_\uparrow
\right|=|F_\uparrow|e^{i\arg F_{\uparrow}}, \end{equation} where
$L_\uparrow$ is the left-hand side of Eq.\ \ref{eq:cwpu} evaluated for
$\arg p_\uparrow=0$. Taking the modulus of Eq.\
\ref{eq:cwp2} we have the general form
\begin{equation}\label{eq:cwpumpmodulus}
|\alpha+|z|^2\beta|=|F_\uparrow|,
\end{equation} where $\alpha$ and $\beta$ are functions of the
intensities of the pump fields. We solve Eq.\ \ref{eq:cwpumpmodulus}
to determine the output intensities that are consistent with the
strength of the pump-up driving, as functions of the intensities of
the pump fields $I_\uparrow(I_\downarrow)$ and $I_\downarrow$. We then
solve the spin-flipped version of Eq.\ \ref{eq:cwpumpmodulus} to
determine the output intensities consistent with the strength of the
pump-down driving. Equating these two intensities gives a nonlinear
equation which we solve to determine $I_\downarrow$ as a function of
the external pumps.

\begin{figure}[t]
\begin{center}
\includegraphics[width=1\linewidth]{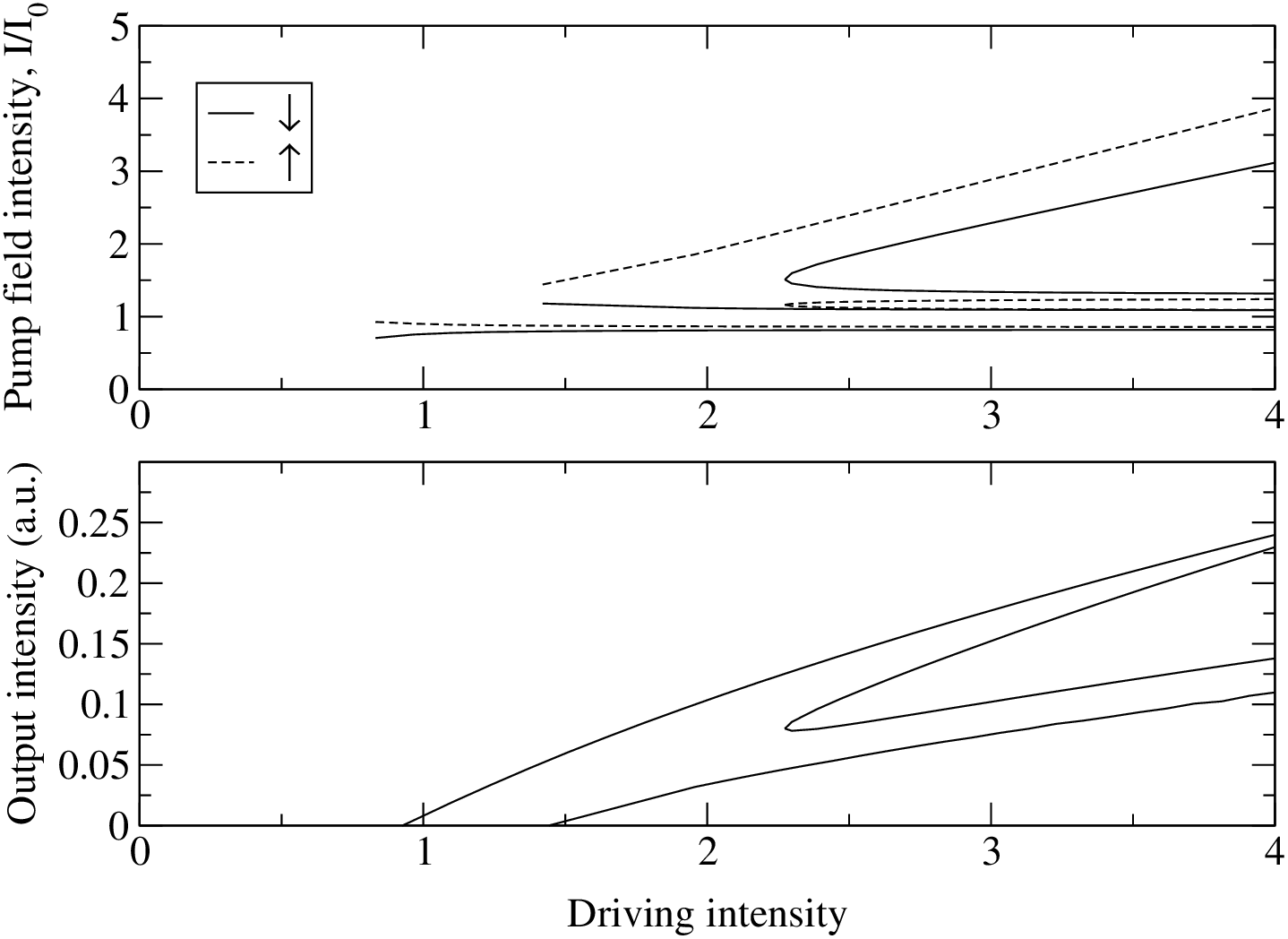}
\end{center}
\caption{Steady-states, for the parameters of Fig.\
  \ref{fig:pumpsolution1}, with an external drive of circularity
  $\sigma=0.1$ and varying intensity. The intensity of the drive is
expressed relative to that of the lowest threshold in the uncoupled
device with the same parameters.}
\label{fig:pumpdepletionsolution1}
\end{figure}

Fig.\ref{fig:pumpdepletionsolution1} illustrates the solution to the
pump equations (\ref{eq:cwpu}). The damping and Hopfield coefficients
for the signal and idler are as in Fig.\ref{fig:pumpsolution1}. For
the pump we have used $\gamma_p=0.1 {\mathrm{meV}}$ and $\xfp^2=0.8$,
which we again estimate to be appropriate to the system reported in
Ref.\ \ \onlinecite{lagoudakis2002}. We have taken an external drive
of fixed circularity, $\sigma=0.1$, and varying intensity. The top
panel shows the intensities of the pump fields, and the bottom panel
shows the output intensity. Increasing the driving intensity from zero
we first find two continuous thresholds, where steady-states appear
starting with zero output intensity. The pump fields at these
thresholds are marked as filled dots on
Fig.\ref{fig:pumpsolution1}. At these points, the pump intensities
match, up to a common factor, with the driving intensities:
$I_{\uparrow\downarrow}=|F_{\uparrow\downarrow}|^2/|\dpp|^2$. For this
small value of $g$, they are approximately the thresholds for each
polarisation of the uncoupled device. With increasing driving, the
output intensity in each of these steady-states increases and, in
constrast to the uncoupled device, the pump fields change. Increasing
the driving intensity still further, we find a third threshold at
which a new steady-state appears, and then splits into two
states. This third threshold is discontinuous, i.e. the solution
appears with a finite output intensity. It corresponds to the pump
fields marked with the open dot on Fig.\ \ref{fig:pumpsolution1}.

\section{Discussion}
\label{section:discussion}

In Ref.\ \onlinecite{tartakovskii2000}, the intensity and
circularity of the signal output was measured under continuous driving
of the pump field. The driving circularity was varied from circular to
linear and the total pump intensity was of the same order of magnitude
as the threshold for a circularly polarised pump. As the pump
circularity $\sigma$ was decreased from one, the signal intensity
increased by a factor of five to a maximum at $\sigma\approx 0.4$, and
then decreased again. The circularity of the signal was approximately
constant, except near to linear pumping, where it dropped to zero.

Considering the pair process underlying the parametric oscillator, one
might expect that without polarisation coupling the output would
monotonically reduce as the pump circularity is reduced: scattering
only occurs within each polarisation, and moving away from circularity
reduces the population of each
polarisation\cite{lagoudakis2002}. However, such an interpretation
overlooks the effects of pump depletion and coherence. Because of
these effects, the steady-state output of a single polarisation pumped
with intensity $I_+$ is actually proportional to
$\sqrt{I_+}-\sqrt{I_{\mathrm{thresh}}}$(Ref.\
\onlinecite{whittaker2001}). For total driving intensities greater
than twice the single-polarisation threshold, there is a critical pump
circularity below which both polarisations are above threshold. Below
this critical circularity, the total output increases as the pump
circularity is reduced, with a local maximum for a linear pump. Thus
an increase in the output as the pump circularity is reduced does not
in general imply the existence of polarisation coupling. However,
without such a coupling it is difficult to explain the dependence of
output on pump circularity reported in Ref.\
\onlinecite{tartakovskii2000}: the enhancement away from circularity
is too strong, and the output peaks for elliptical, rather than
circular or linear, pumping.

The lower branches shown on Fig.\ref{fig:pumpsolution2} have no
structure to suggest that they would give a maximum output for an
elliptically polarised pump. However, the upper branches illustrated
for $g=5$ and $g=7$ do have such structure. As the pump circularity is
reduced from one, we would first go up through the threshold for these
solutions, then move back towards it, and for some parameters go below
it again as we approach linear pumping. The circularity of the turning
points illustrated for $g=7$ is $\sigma=0.47$. While this is roughly
consistent with the experimental results, a detailed fit to the
experiment is beyond the scope of the present paper. It would involve
a large number of parameters, and we expect the results to be
sensitive to details left out of the present model such as the
blueshifts.

It is only stable steady-states which are relevant to continuously
pumped experiments. We have analysed the stability of some of the
steady-state solutions to our model for a special case in which all
the effective damping rates are equal,
$\gamma_p^\prime=\gamma_s^\prime=\gamma_i^\prime$. For $g=0.1$, there
are two solutions with continuous thresholds and two with
discontinuous thresholds, as there are in Fig.\
\ref{fig:pumpdepletionsolution1}. We find that only the solution with
the lowest threshold is stable. For $g=2.0$, we find only one solution
with a continuous threshold, as well as two with discontinuous
thresholds. In that case, both the continuous solution and one of the
discontinuous solutions is stable. Thus it is possible for our model
to have stable solutions other than that with the lowest threshold,
and to have more than one stable solution.

The pulsed experiments of Ref.\ \onlinecite{lagoudakis2002} have
recently been addressed by Kavokin et al.\ \cite{kavokin2003}. Their
theory reproduces the experimentally observed rotations of the linear
polarisation without scattering between polarisations. In their
theory, the rotation of the linear polarisation comes from the
different blueshifts of the two circular polarisation states. The
present work does not include this effect, because we have taken the
detunings of the two polarisation states to be the same. We expect
that if the blueshifts are small compared with the interactions
between polarisations then the steady-states will only contain a
single frequency for the pump, signal and idler, and the treatment
given here will be qualitatively correct. However, a splitting of the
two circular polarisation states might be an alternative explanation
for the steady-states results of Ref.\ \onlinecite{tartakovskii2000}.

The agreement between the polarisation rotations seen in the pulsed
experiments\cite{lagoudakis2002} and the theory of Ref.\
\onlinecite{kavokin2003} suggests that a polarisation coupling term is
not required to explain these results. However, this does not imply
that the polarisation coupling is always irrelevant. The polarisation
rotations produced by a coupling could be small compared with those
produced by the blueshifts, allowing a good fit to this aspect of the
data without a coupling. Note also that the pulsed experiments are
done at much higher excitation powers, typically around a hundred
times greater, than the steady-state experiments. The polarisation
coupling could also depend on sample parameters such as the energy
difference between the pump polaritons and the
biexciton\cite{tartakovskii2000}.

An interaction of the form we propose corresponds microscopically to
an interaction between excitons of opposite spin, which could come
from the Coulomb interactions between the electrons and the holes in
the excitons. To first order in these Coulomb interactions, the
interaction between opposite spin excitons is
negligible\cite{ciuti1998,maialle1994}, because in that approximation
small wavevector scattering is dominated by electron-electron and
hole-hole exchange. Thus a finite value of $g$ implies the
significance of higher-order processes, which can produce interactions
between opposite spin excitons\cite{maialle1994}. We suggest that the
higher-order processes could involve the $m_j=\pm 2$ excitons, which
are close in energy, and produce an interaction at second
order. Another recent suggestion\cite{inoue2000} is that the
interaction involves excited states of the excitons.

\section{Conclusions}
\label{section:conclusions}

We have studied the steady-states of a model of microcavity parametric
oscillation with coupled polarisations. For small values of the
coupling, we find two steady-states corresponding to those of the
uncoupled device. However, due to the increased scope for arranging
the pump depletion, there are also two steady-states which appear
discontinuously, i.e. with a finite value of the output intensity, as
the driving intensity is increased. For general values of the coupling
Fig.\ref{fig:pumpsolution2} suggests that there will be either
one or two continuous solutions, depending on the coupling and pump
circularity. There may also be discontinuous solutions. For some
parameters more than one steady-state can be stable, in which case it
should be possible to observe switching between the states induced
either by noise or by external probes.

The coupling between polarisations introduces two types of mixing term
into the equations of motion for the fields. As well as the
straightforward analog of the process considered in the uncoupled
model, there are processes which exchange the spins of two
fields. Such spin-flip processes lead to a rotation of the output
polarisation with respect to the pump. They also produce shifts in the
output frequencies with varying pump intensity or circularity, even in
the absence of the shifts associated with the mean-field
exciton-exciton interaction.

\begin{acknowledgments}

We thank Jeremy Baumberg and Peter Littlewood for discussions of this
work. PRE acknowledges the support of a research fellowship from
Sidney Sussex College, Cambridge, and DMW that of an advanced research
fellowship from the EPSRC(GR/A11601). This work is also supported by the EU
network ``Photon mediated phenomena in semiconductor nanostructures''.

\end{acknowledgments}

\end{document}